\documentclass[%
 reprint,
 amsmath,amssymb,
 aps,
 pra,
floatfix,
]{revtex4-2}

\pdfoutput=1
\usepackage{graphicx}
\usepackage{dcolumn}
\usepackage{bm}
\usepackage{color, soul}
\DeclareMathOperator{\Tr}{Tr}

\begin{document}

\preprint{APS/123-QED}

\title{Stimulated Slowing of Yb Atoms on the Narrow \textsuperscript{1}S\textsubscript{0} $\rightarrow$ \textsuperscript{3}P\textsubscript{1} Transition}

\author{Tanaporn Na Narong}
\email{tn282@stanford.edu}
\author{TianMin Liu}%
\author{Nikhil Raghuram}%
\altaffiliation{Department of Physics, Virginia Polytechnic Institute and State University, Blacksburg, VA 24061}%
\author{Leo Hollberg}%
\email{leoh@stanford.edu}
\affiliation{%
Hansen Experimental Physics Laboratory, Department of Physics, Stanford University, Stanford, CA 94305
}%

\date{\today}

\begin{abstract}
We  analyzed bichromatic and polychromatic stimulated forces for laser cooling and trapping of Yb atoms using only the narrow \textsuperscript{1}S\textsubscript{0} $\rightarrow$ \textsuperscript{3}P\textsubscript{1} transition. Our model is based on numerical solutions of optical Bloch equations for two-level atoms driven by multiple time-dependent fields combined with Monte-Carlo simulations, which account for realistic experimental conditions such as atomic beam divergence, geometry, and Gaussian laser modes. Using 1 W of laser power, we predict a loading rate of $\approx$ 10\textsuperscript{8} atoms/s into a 556 nm MOT with a slowing force of $\approx 60F_{rad}$. We show that a square wave modulation can produce similar stimulated forces with almost twice the velocity range and improve the MOT loading rate of Yb atoms by up to 70\%.

\end{abstract}

\maketitle


\section{\label{sec:intro}Introduction}
Cold atom systems and quantum sensors rely upon sources of low velocity atoms with well-controlled motional degrees of freedom. Magneto-optical traps (MOTs) are the starting point of many experiments including atomic clocks, atom interferometry, optical lattices, tweezers, fountains, and quantum degenerate gases. In this paper we analyze a promising method to reduce the complexity of cold atom sources for Yb while maintaining good atom numbers and low temperatures. Aspects of the approach may be applicable to other atomic systems.

\begin{figure}[ht]
\includegraphics[width=0.46\textwidth]{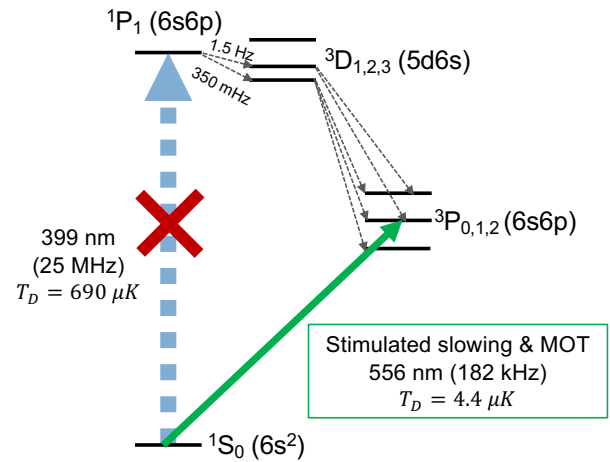}
\caption{\label{fig:energy_levels}Energy level diagram of Yb showing the \textsuperscript{1}S\textsubscript{0} $\rightarrow$ \textsuperscript{3}P\textsubscript{1} transition (solid arrow), which is used for both stimulated slowing and magneto-optical trapping. The stimulated slowing method eliminates the need for the traditional first-stage cooling on the \textsuperscript{1}P\textsubscript{1} transition and helps avoid population loss via decays from \textsuperscript{1}P\textsubscript{1} to the lower-lying triplet D-states, \textsuperscript{3}D\textsubscript{2,1} (5d6s), which can lead to further decays to dark \textsuperscript{3}P\textsubscript{2,0} states where atoms are lost from the cooling cycles or MOTs without repumping.}
\end{figure}

Many approaches have been developed to create better cold atom sources with goals of increasing cold atom flux, MOT loading rates, atom densities, and/or decreasing temperatures. Laser cooling is often done in two stages for closed-shell, alkaline-earth-like atoms (e.g. Ca, Sr, Yb), which are of current interest for optical atomic clocks \cite{poli2013optical}, quantum gases \cite{sugawa2011bose, hara2011quantum}, quantum measurements \cite{covey2019telecom}, and atom interferometry \cite{rudolph2020large, hu2019sr}. The first stage cools on a broad transition to produce fast momentum transfers and large slowing forces. The second stage cools on a narrow transition to reach lower temperatures (in $\mu$K for Yb and Sr). The energy level diagram for Yb in Fig.\ref{fig:energy_levels} shows the first-stage cooling transition \textsuperscript{1}S\textsubscript{0} $\rightarrow$ \textsuperscript{1}P\textsubscript{1} at 399 nm and the second-stage cooling transition  \textsuperscript{1}S\textsubscript{0} $\rightarrow$ \textsuperscript{3}P\textsubscript{1} at 556 nm.

Previous laser cooling methods developed to improve loading efficiency for MOTs that trap Yb and other alkaline-earth-like atoms include tailored Zeeman slowers \cite{loftus2001laser, loftus_patent_2014, wodey2021robust} and pre-loading 2D-MOTs \cite{dorscher2013creation}. Other methods include adiabatic rapid passage \cite{stack2011numerical, norcia2018narrow, petersen2019sawtooth, bartolotta2020speeding, malinovskaya2021laser}, Sisyphus-like deceleration \cite{chen2019sisyphus}, two-photon cooling \cite{magno2003two}, two-color cooling \cite{lee2015core, plotkin2020crossed}, and quenched narrow-line cooling \cite{curtis2001quenched, grain2007feasibility} that takes advantage of the narrower transitions to achieve very low temperatures. Combining several of these techniques can produce an ultracold atom source with higher phase-space densities \cite{chen2019continuous}. With these approaches, experimental systems usually require multiple lasers: a first-stage cooling laser, a second-stage cooling laser, and sometimes repumping lasers to retrieve atoms from dark states. As shown in Fig.\ref{fig:energy_levels}, Yb atoms can decay from the \textsuperscript{1}P\textsubscript{1} state to the lower-lying triplet D-states, \textsuperscript{3}D\textsubscript{2,1}, and may end up in the dark \textsuperscript{3}P\textsubscript{2,0} states. Removing the first-stage cooling will avoid the dark-state losses and allow for a more compact system with only one cooling laser.

\begin{figure}[ht]
\includegraphics[width=0.48\textwidth]{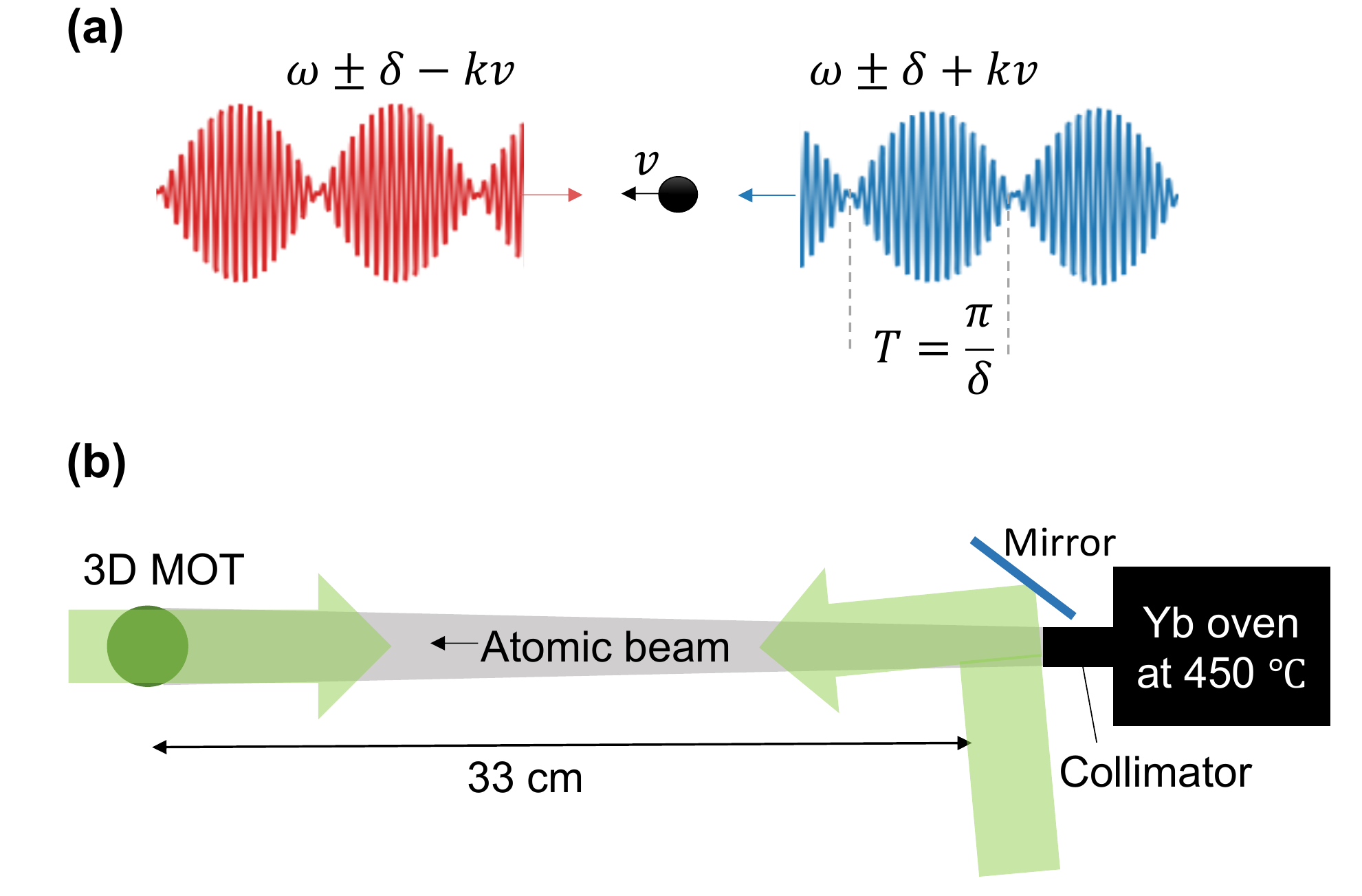}
\caption{\label{fig:setup}(a) An atom traveling in counter-propagating bichromatic light fields with a velocity $v$. The laser beam frequencies are detuned by $+kv$ and $-kv$ to compensate for Doppler shifts. (b) A schematic diagram of our experimental setup showing a Yb atomic beam leaving an oven at 450 $^\circ$C and two counter-propagating laser beams at 556 nm. The atoms are slowed over a distance of 33 cm before reaching the MOT region. Our current Yb atomic beam is partially collimated by a tubular array nozzle and diverges with a half-angle $\approx$ 17 mrad.}
\end{figure}

Our goal is to eliminate the first-stage cooling on the \textsuperscript{1}S\textsubscript{0} $\rightarrow$ \textsuperscript{1}P\textsubscript{1} transition and only use the narrow \textsuperscript{1}S\textsubscript{0} $\rightarrow$ \textsuperscript{3}P\textsubscript{1} transition for both slowing and trapping of Yb atoms while maintaining a high loading rate into the MOT. One limitation of laser cooling on this narrow transition is that the spontaneous optical forces are limited by the relatively small linewidth $\gamma = 2\pi \times 182$ kHz. To overcome this limit, driving stimulated transitions in atoms at much higher rates than the spontaneous decay rate can produce a slowing force $F \gg F_{rad} = \hbar k \gamma/2$. We theoretically demonstrate in this paper that using stimulated methods on the narrow \textsuperscript{1}S\textsubscript{0} $\rightarrow$ \textsuperscript{3}P\textsubscript{1}  transition in Yb can produce large forces and a high loading rate into the MOT.

Stimulated slowing methods with atomic beams \cite{metcalf_colloquium_2017} have been demonstrated on alkali atoms \cite{voitsekhovich1991stimulated, grimm1990observation, sodingShortDistanceAtomicBeam1997, williams_measurement_1999, williams_bichromatic_2000, liebischAtomnumberAmplificationMagnetooptical2012} and metastable helium \cite{cashen_strong_2001, partlow2004bichromatic, chiedaBichromaticSlowingMetastable2012} but not on alkaline-earth-like atoms. Recent efforts have extended the concepts \cite{chiedaProspectsRapidDeceleration2011, galica_four-color_2013, dai2015efficient, ilinova2015stimulated, aldridge_simulations_2016, yin2018optically, wenz_large_2020} and experiments \cite{kozyryev2018coherent, galica2018deflection} for both atoms and  molecules. Stimulated emissions can be driven by ultrashort $\pi$-pulses \cite{goepfert1997stimulated, ilinova2011doppler, long2019suppressed} or bichromatic CW beams. Following the early observations on Na \cite{voitsekhovich1991stimulated, grimm1990observation}, Söding et al. decelerated Cs atoms with a stimulated force $\approx 10F_{rad}$ using counter-propagating, bichromatic light fields with symmetric detunings $\pm \delta$ \cite{sodingShortDistanceAtomicBeam1997}. That same setup was used in the first part of our analysis, as shown in Fig.\ref{fig:setup}a. Söding et al. also developed computer simulations to calculate the stimulated forces by solving the optical Bloch equations (OBEs) for two-level systems. Those simulations were adopted and applied to later bichromatic slowing experiments by Metcalf and his colleagues. They studied and measured the bichromatic force (BCF) in more detail on Rb \cite{williams_measurement_1999, williams_bichromatic_2000} and He* \cite{cashen_strong_2001, partlow2004bichromatic} atoms. They also developed a theoretical model of BCF using a dressed-atom picture \cite{yatsenkoDressedatomDescriptionBichromatic2004}, which predicted the velocity range of the BCF more accurately than the intuitive $\pi$-pulse interpretation, which treats the bichromatic waves as individual non-overlapping pulses. More recent experiments optimized bichromatic slowing further to enhance the Rb MOT loading rate \cite{liebischAtomnumberAmplificationMagnetooptical2012} and extend the velocity range of BCF to over 300 m/s using laser frequency chirp \cite{chiedaBichromaticSlowingMetastable2012}. We discuss in Sec.\ref{subsec:chirp} how we applied the chirped BCF method on Yb for efficient MOT loading.

We developed computer simulations of stimulated slowing of Yb atoms using only the narrow 556 nm transition for both slowing and trapping with a MOT. We first describe our model in Sec.\ref{sec:methods}. Because the \textsuperscript{1}S\textsubscript{0} $\rightarrow$ \textsuperscript{3}P\textsubscript{1} transition in Yb is nearly closed, a two-level model is well suited to describe the system. We developed Monte Carlo simulations to predict the MOT loading rate and inform the experimental design for our existing AOSense atomic beam system (Fig.\ref{fig:setup}b). Section \ref{subsec:chirp} outlines the chirped BCF method \cite{chiedaBichromaticSlowingMetastable2012} and how we optimized the MOT loading rate by choosing experimental parameters such as bichromatic detuning $\delta$, slowing beam diameters, and beam intensities, under physical constraints including the Gaussian beam profile, gravity, and limited optical power. Using 1 W of laser power, the chirped BCF method using only the 556 nm laser can achieve a loading rate of $10^{8}$ atoms/s, which is comparable to the loading rate into a 556 nm MOT achieved by a 399 nm Zeeman slower \cite{guttridge_direct_2016}. We analyze other approaches to improve the stimulated forces and the MOT loading rate in the later sections.

Our analysis extends beyond the bichromatic force to polychromatic forces in Sec.\ref{subsec:square}. Inspired by the four-color stimulated forces presented in \cite{galica_four-color_2013}, we demonstrate that adding higher odd harmonics of the bichromatic light via a square wave amplitude modulation roughly doubles the velocity range of the force and improves the MOT loading rate by 70\% compared to the chirped BCF method with the same total laser power. We also discuss the time evolution of the two-level-atom in the square wave amplitude-modulated light fields using Bloch vector trajectories. We show in Sec.\ref{subsec:phasemod} that a square wave phase modulation produced similar stimulated forces, which opens up a possibility of utilizing phase modulators in stimulated slowing experiments. To broaden the velocity range of the force, we explored a broadband cooling approach that superposes two BCF profiles at two different detunings. Although our calculations show that interference makes BCF vanish for a single two-level system \cite{partlow2004bichromatic}, broadband cooling may be feasible in molecules and more complex multi-level systems \cite{wenz_large_2020}. Overall, our simulation results provide insights into bichromatic and square-wave stimulated forces on two-level atoms and inform the experimental design that optimizes the loading rate of Yb atoms into the MOT using only the 556 nm laser.

\section{\label{sec:methods}Numerical model}
To investigate different stimulated slowing approaches, we developed our initial model in Python to calculate the bichromatic force (BCF) on two-level atoms. Similar to what was implemented in \cite{sodingShortDistanceAtomicBeam1997}, our initial model assumes a two-level atom traveling in a bichromatic standing wave. Each amplitude-modulated wave consists of two beating frequencies $\omega \pm \delta$, which are detuned to compensate for the Doppler shifts (Fig.\ref{fig:setup}a). Other key parameters include the Rabi frequency $\Omega$ and relative phase $\chi$ between the blue-detuned and red-detuned light fields.  

The force on an atom is calculated numerically from the density matrix, $\rho(t)$, and the two-level atom's Hamiltonian, $H(t)$. We applied the rotating wave and dipole approximations on the 2 $\times$ 2 Hamiltonian matrix. When the bichromatic light fields are detuned by $\pm \delta$ from resonance, the diagonal terms $H_{00}$ and $H_{11}$ of $H(t)$ are zero. The off-diagonal components $H_{01} = H_{10}^\dagger$ are given by
\begin{equation}
    \begin{split}{\label{eqn:H_BCF}}
    H_{01}(z, t) &= \frac{\Omega}{2}\left[ e^{-ikz}(e^{i\delta t + \chi} + e^{-i\delta t - \chi}) + e^{ikz}(e^{i\delta t} + e^{-i\delta t})\right] \\
    &= \Omega \left[ e^{-ikz}\cos{(\delta t + \chi)} + e^{ikz}\cos{(\delta t)} \right]
\end{split}
\end{equation}

\noindent
where the Rabi frequency $\Omega = \frac{eE_0}{\hslash}\langle 1 \lvert r \rvert 0 \rangle$ and $E_0$ is the electric field amplitude of each of the four CW beams.\\

We then solve  the optical Bloch equations (OBEs) for the density matrix $\rho$ using a built-in Linblad Master equation solver from the QuTip library \cite{johanssonQuTiPOpensourcePython2012, johanssonQuTiPPythonFramework2013}. The Linblad master equation (Eq.\eqref{eqn:master}) includes spontaneous emissions via a collapse operator $C = \sqrt{\gamma}\lvert 0 \rangle \langle 1 \rvert$, where $\lvert 0 \rangle$ and $\lvert 1 \rangle$ denote a two-level atom's ground and excited states, respectively. 

\begin{equation}{\label{eqn:master}}
    \Dot{\rho}(t)=-i \hslash [H(t), \rho(t)] + \frac{1}{2} [2 C \rho(t) C^{\dagger} - \rho(t)C^{\dagger}C - C^{\dagger}C\rho(t)]
\end{equation}

\begin{figure}[ht]
\centering
\includegraphics[width=0.49\textwidth]{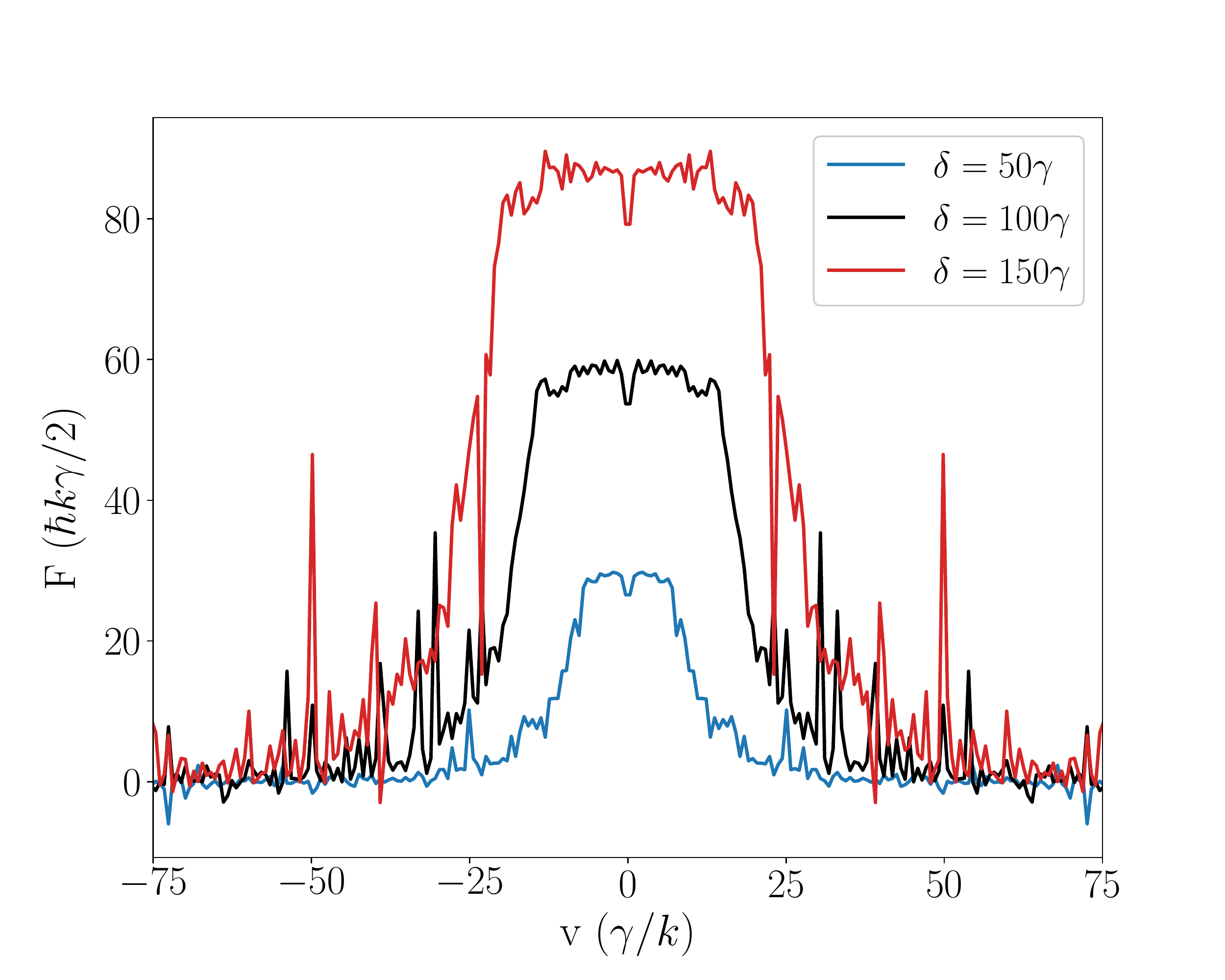}
\caption{\label{fig:forcevelocity} Bichromatic force (BCF) as a function of atomic velocity for $\delta = 50\gamma, 100\gamma, 150\gamma$, Rabi frequency $\Omega = \sqrt{3/2}\delta$, $\chi = \pi/4$. The magnitude and range of the force scale linearly with $\delta$. At $\delta = 150\gamma$ and $\Omega = 184\gamma$, the maximum stimulated force is near $90F_{rad}$ and acts over the velocity range of $\pm30\gamma/k \approx 3$ m/s for the $^3 P_1$ transition in Yb.}
\end{figure}

\noindent
After solving Eq.\eqref{eqn:master} for $\rho$, our program calculates the force from $F(t)= \hslash \frac{\partial}{\partial z}\Tr{(\rho H)}$. We found that setting $\Omega/\delta = \sqrt{3/2}$ and $\chi = \pi/4$ optimized the magnitude and velocity range of the BCF. Our results are consistent with previous numerical work \cite{sodingShortDistanceAtomicBeam1997, yatsenkoDressedatomDescriptionBichromatic2004, galica_four-color_2013}. Figure \ref{fig:forcevelocity} shows the stimulated forces plotted against atomic velocity for three different values of $\delta$ and the respective Rabi frequency $\Omega = \sqrt{3/2}\delta$. At this optimal ratio and phase $\chi = \pi/4$, the magnitude of the bichromatic force is given by
 
\begin{equation}{\label{eqn:BCF}}
F_{BCF} = \frac{\hslash k \delta}{\pi} 
\end{equation} .

\noindent
Since we can choose $\delta \gg \gamma$, $F_{BCF}$ can be much larger than the radiative force $F_{rad} = \hslash k \gamma / 2$ and have a broad velocity range of $\Delta v \sim \delta/2k \gg \gamma/k$ \cite{yatsenkoDressedatomDescriptionBichromatic2004}. The red curve in Fig.\ref{fig:forcevelocity} corresponds to $\delta = 150 \gamma$ and $\Omega = 184\gamma \approx 2\pi \times 33$ MHz, which produces a BCF with magnitude $\approx 90F_{rad}$. Another key feature of the BCF is sharp Doppleron resonances at velocities $v = \pm\delta/(2n + 1)k, n = 0, 1, 2, ...$ \cite{minogin1979resonant}. Nevertheless, these sudden increases in the stimulated force were not observed in bichromatic force measurements \cite{williams_measurement_1999, williams_bichromatic_2000} and do not affect the slowing process significantly.

Both the magnitude and velocity range of the bichromatic force show promise for stimulated slowing in loading atoms into a MOT \cite{liebischAtomnumberAmplificationMagnetooptical2012}. We estimate the MOT loading efficiency for Yb atoms using Monte Carlo simulations with roughly 20,000 atoms. For an individual atom with a given velocity and position, we compute the bichromatic force by following the above procedure and then integrate the force to update the atom's velocity and position at every 10 $\mu$s interval. This time interval is approximately the time it takes for an atom's velocity to change by $\gamma /k$ under the stimulated force. 

Our simulations account for realistic experimental conditions as illustrated in Fig.\ref{fig:setup}b. Yb atoms leave a 450 $^{\circ}$C oven at a constant flux and travel in a bichromatic light field, defined by the key parameters $\delta$ and Rabi frequency $\Omega$, towards a 3D 556 nm MOT located 33 cm away from the oven aperture that has a small diameter of 4 mm. The atomic beam is slowly diverging with 17 mrad half-angle divergence. We take into account the laser's Gaussian beam profile, which causes a spatial variation in laser intensity and hence Rabi frequency $\Omega$. Our model includes the effects of gravity. Monte Carlo simulations enabled us to estimate the fraction of atoms that are slowed to the MOT's capture velocity and the respective loading rate. Assuming a 2-cm MOT beam diameter, Yb atoms would need to be slowed from several 100 m/s to the MOT capture velocity of only $\approx$5 m/s.

We have estimated the effects of momentum diffusion due to spontaneous emission. Due to the relatively long lifetime of the \textsuperscript{3}P\textsubscript{1} state (874 ns), Monte Carlo simulations estimate that an atom in resonance with the bichromatic light fields undergoes up to 49,000 stimulated and 1600 spontaneous emissions in the slowing duration of 3 ms. With the recoil velocity of 4 mm/s/photon, we estimate the random diffusion velocity to be $\approx$16 cm/s, which is negligible for this system. These effects are even smaller when atoms are driven by square wave amplitude-modulated light (Sec.\ref{subsec:square}). In that case, the number of spontaneous emissions is reduced by 33\% and the estimate random diffusion velocity decreases to 13 cm/s. The heating effects from momentum diffusion would be more prominent for atoms with shorter lifetimes. For Yb, we neglected these effects and focused our efforts on achieving a large velocity slowing range (e.g. 200 m/s) and optimizing the loading rate into the MOT. We discuss several approaches as follows.

\section{\label{sec:results}Results \& Discussion}

\subsection{\label{subsec:chirp}Chirped BCF}
Chieda and Eyler demonstrated on He* atoms that laser frequency chirp can extend the velocity range of stimulated slowing without requiring a very large bichromatic detuning $\delta$ \cite{chiedaBichromaticSlowingMetastable2012}. The laser frequency was swept in a sawtooth manner to keep the laser detuning in resonance with the moving atoms while they are being slowed. This method can extend the slowing velocity range from a few 10 m/s to $>$200 m/s without requiring larger $\delta$ and Rabi frequency $\Omega$. Although the time dependence of the chirp method does not capture all of the atoms, the method does reduce the power requirement and avoids the large $\delta$ regime where the bichromatic force reportedly vanishes \cite{chiedaBichromaticSlowingMetastable2012}.

\begin{figure}[ht]
\includegraphics[width=0.49\textwidth]{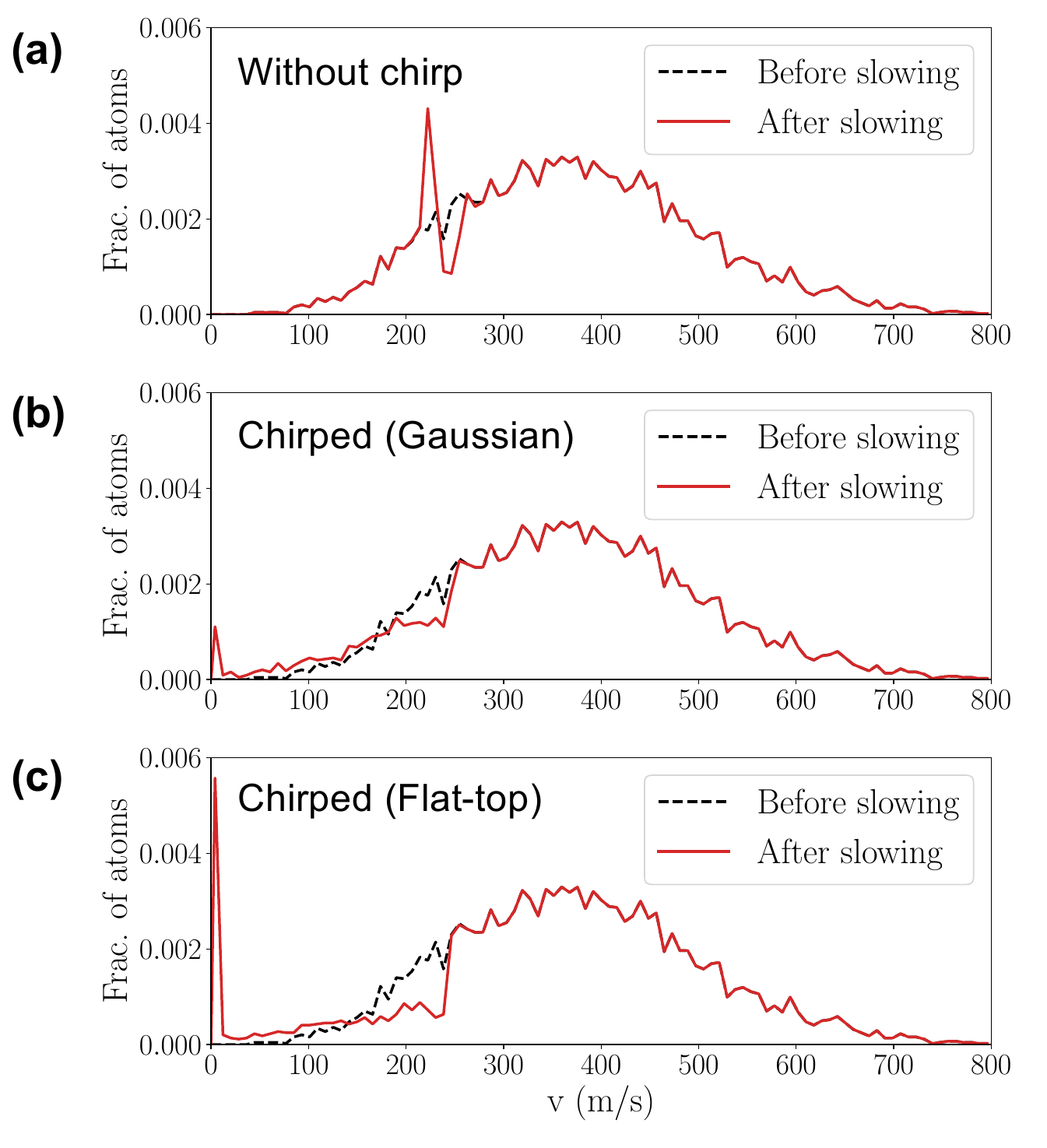}
\caption{\label{fig:vdist} Velocity distributions before (dashed) and after (solid) slowing with (a) a fixed detuning at 240 m/s and (b) chirped detuning from 240 to 10 m/s in 2 ms. Gaussian beam diameter of 8 mm, $\delta = 200 \gamma$ and peak Rabi frequency $\Omega = 244 \gamma$ for both (a) and (b). (c) Same chirp parameters and laser power as (b) but with a flat-top beam profile of the same diameter. Rabi frequency $\Omega = 173 \gamma$ is constant across the beam profile and $\delta$ was adjusted to $141\gamma$.}
\end{figure}

Our Monte Carlo simulation results in Fig.\ref{fig:vdist} show that laser frequency chirp can significantly increase the slowing range and fraction of slow atoms below the MOT capture velocity 5 m/s. In Fig.\ref{fig:vdist}a, the laser detuning was fixed at 240 m/s. Even at $\delta=200\gamma$ and $\Omega = 244\gamma$, the affected atoms were only slowed by a few 10 m/s range, and not surprisingly the number of atoms near zero velocity remained unchanged. In contrast, sweeping the detuning from 240 to 10 m/s in 2 ms substantially broadened the slowing velocity range and increased the population of atoms with velocities below the capture velocity of 5 m/s by at least four orders of magnitude. This result demonstrates that the laser frequency chirp method is an effective method in loading the MOT. Our simulations also enabled us to investigate how other experimental conditions, such as the laser beam profile, affect the MOT loading rate. 

As Fig.\ref{fig:vdist}b and \ref{fig:vdist}c show, with stimulated slowing the realistic laser Gaussian beam profile reduces the MOT loading rate relative to an unrealistic uniform intensity profile. Because the stimulated force is strongly dependent on the $\Omega/\delta$ ratio and the Rabi frequency $\Omega$ rapidly decays with the atoms’ distance from the beam axis, only atoms near the center of the beam experience a significant force and are slowed to the target capture velocity. In this particular example, the MOT loading rate was roughly five times smaller when using the Gaussian beam profile (Fig.\ref{fig:vdist}b) compared to the flat-top beam profile (Fig.\ref{fig:vdist}c), in which Rabi frequency $\Omega$ remains constant and optimized across the beam area. Both simulations used the same chirped detuning from 240 to 10 m/s, the same laser power, and the same beam diameter of 8 mm, which was twice as large as the oven aperture diameter of 4 mm, to compensate for the atomic beam divergence and maximize the number of atoms slowed. In addition to the laser power, geometric constraints of the experimental hardware must be taken into account to optimize the loading rate into the MOT.

To optimize the slowing force on atoms and the MOT loading rate, we kept the $\delta / \Omega$ ratio at $\sqrt{3/2}$ at the center of the Gaussian beam. The Rabi frequency $\Omega$ is proportional to the square root of the laser intensity. Therefore, $\delta$, $\Omega$, and hence the stimulated force, are limited by laser intensity. For a fixed laser power, there is a trade-off between laser intensity, which sets the magnitude of the slowing force, and laser beam size, which determines the area across the laser beam profile where the force is present. Chirp parameters including the chirped detuning range and chirp period also play a role. A high laser intensity and large force allow a higher chirp rate, which means the starting detuning can be set to a higher velocity to slow a larger fraction of atoms that follow the effusion velocity distribution $f(v) \propto v^3 \exp(-mv^2/k_B T)$. Therefore, the fraction of atoms in the atomic beam that will be slowed is strongly dependent on the choice of chirp parameters as well as $\delta$, $\Omega$ and the laser Gaussian beam diameter.  

\begin{figure}[b]
\includegraphics[width=0.48\textwidth]{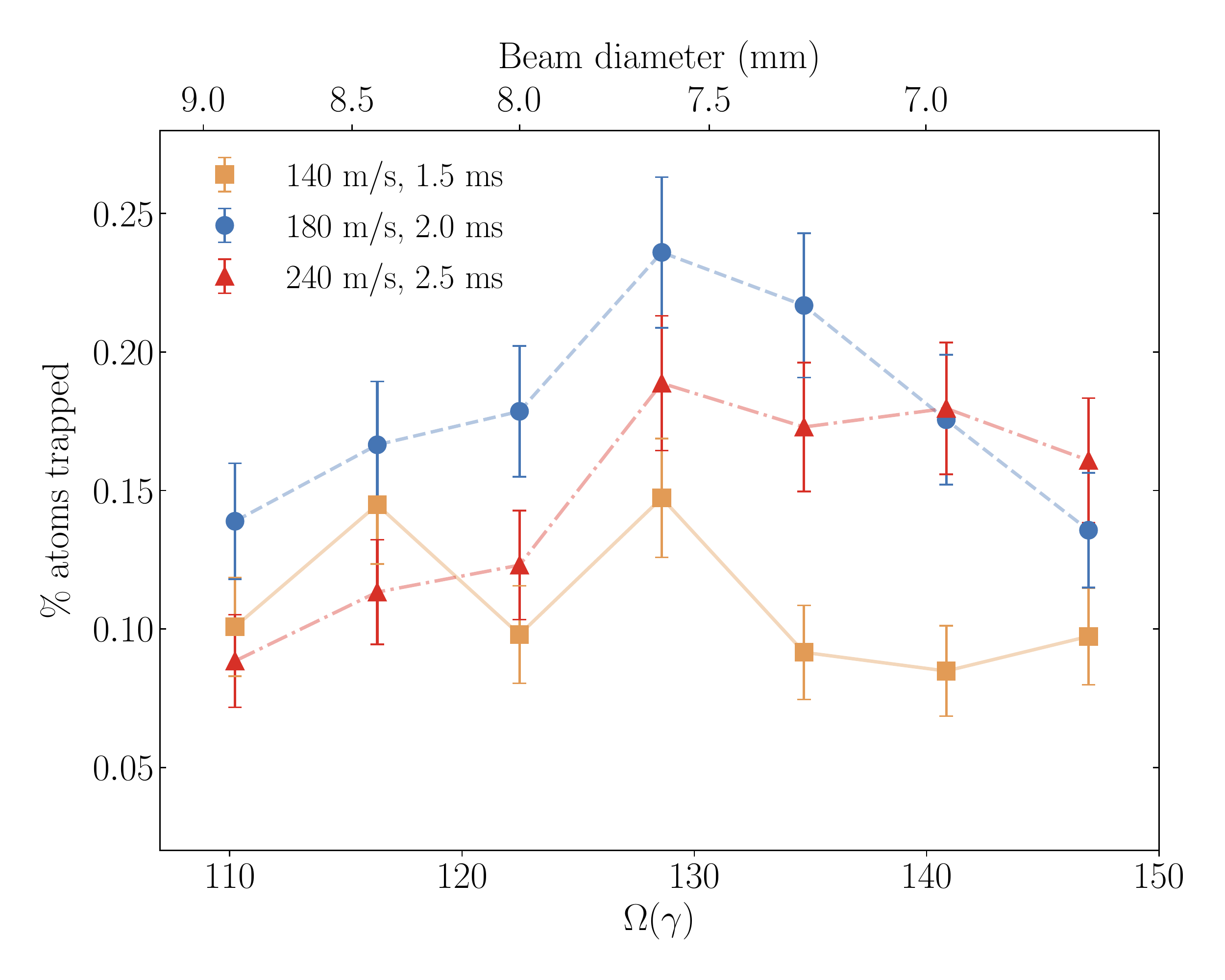}
\caption{\label{fig:loadingrate_BCF} Percentage of atoms in the atomic beam that can be trapped by the 556 nm MOT after bichromatic slowing plotted against Rabi frequency $\Omega$ and laser beam diameter for three different chirp settings: 140 $\rightarrow$ 10 m/s in 1.5 ms (square, solid line), 180 $\rightarrow$ 10 m/s in 2 ms (circle, dashed line), and 240 $\rightarrow$ 10 m/s in 2.5 ms (triangle, dot-dashed line). The error bars represent 95\% confidence interval. Up to 0.24\% of atoms can be trapped using $\delta = 105\gamma$, $\Omega = 129\gamma$ with a Gaussian beam 1/e diameter of 7.6 mm and a laser detuning chirp from 180 to 10 m/s in 2 ms.}
\end{figure}

Under realistic experimental constraints including the Gaussian beam profile and total laser power of 1 W distributed equally between the stimulated slowing beams, we determined $\delta, \Omega$, and chirp parameters that optimized the loading rate of atoms into a 556 nm MOT. Figure \ref{fig:loadingrate_BCF} shows some of the best results predicted by our Monte Carlo simulations of over 20,000 atoms, using three different chirp settings: 140 $\rightarrow$ 10 m/s in 1.5 ms (square, solid line), 180 $\rightarrow$ 10 m/s in 2 ms (circle, dashed line), and 240 $\rightarrow$ 10 m/s in 2.5 ms (triangle, dot-dashed line). The error bars shown on the plot represent the 95\% binomial confidence interval. We found that the optimal settings for bichromatic slowing of Yb for the optical power limit of 1 W were $\delta = 105\gamma$, $\Omega = 129\gamma$ with a Gaussian beam 1/e diameter of 7.6 mm and a laser detuning chirp from 180 to 10 m/s in 2 ms. With these settings, up to 0.24$\pm$0.03\% of the total atomic flux from the oven can be trapped by the 556 nm MOT with 2 cm diameter located 33 cm away from the oven aperture. Here we assume the MOT beams are retroreflected and have a peak intensity of about 20 times the saturation intensity of the 556 nm transition, corresponding to a total power of $<$50 mW for the MOT. 

Assuming an atomic flux of 10\textsuperscript{11} atoms/s leaving the oven (as observed experimentally), the predicted loading rate of 0.24\% is equivalent to $2.4 \times 10^8$ atoms/s. This MOT loading rate using the chirped bichromatic slowing method is comparable to the 556 nm MOT loading rate achieved using a conventional 399 nm Zeeman slower \cite{guttridge_direct_2016}.  We were encouraged by results by Kawasaki et al. who showed that loading Yb directly into a 556 nm MOT without first-stage cooling was possible, but the loading rates were low (under 10\textsuperscript{4} atoms per second) \cite{kawasaki_two-color_2015}. Our simulation results so far have shown that bichromatic slowing with the 556 nm laser may be able to replace a 399 nm Zeeman slower in loading Yb atoms into the MOT. We also explore different approaches to improve this stimulated slowing method. We show later in Sec.\ref{subsec:square} that a square wave modulation method can improve the MOT loading rate even further.

\subsection{\label{subsec:square}Square wave amplitude modulation}
Although the bichromatic force is already an effective method for stimulated slowing of Yb atoms, this force can be enhanced by adding higher harmonics of the bichromatic light \cite{galica_four-color_2013}. Galica et al. presented numerical calculations of a four-color polychromatic force \cite{galica_four-color_2013}, which has a larger magnitude and broader velocity range than BCF, and is produced by adding the third harmonic of the bichromatic detuning $\delta$ to make the pulses narrower in time and more similar to separate $\pi$-pulses. Their results motivated us to investigate the use of square wave modulation and the effects of adding higher harmonics on stimulated forces on Yb.

Starting from the BCF model we developed, we modified the Hamiltonian to describe square amplitude-modulated light. The cosine amplitude modulation terms in Eq.\eqref{eqn:H_BCF} are replaced by a square wave modulation at frequency $\delta_{sq}$. The light fields now consist of two counter-propagating beams, each with Rabi frequency $\Omega$, a square amplitude modulation at frequency $\delta_{sq}$, and a phase difference $\chi$ between each beam. We investigated both near-ideal square waves produced by a built-in square wave function in the \texttt{Scipy} signal processing library and approximations from a truncated Fourier series of a square wave. The Fourier series that contains $n$ harmonics consists of odd harmonics $\pm\delta_{sq}, \pm3\delta_{sq}, \pm5\delta_{sq}, ..., \pm(2n - 1)\delta_{sq}$ with decreasing amplitudes. For example, a square wave with only two harmonics ($n = 2$) has frequency components $\pm\delta_{sq}$ and $\pm3\delta_{sq}$ with Rabi frequencies $\Omega$ and $\Omega/3$, respectively. One limitation of the square wave modulation is that we cannot specify the amplitudes or phases of different harmonics independently. We show later that this is not a problem provided we find the right $\Omega$ and $\chi$. The off-diagonal element $H_{01}^{sq}$ of the new Hamiltonian now contains the square wave Fourier series, and is given by

\begin{equation}
\begin{split}{\label{eqn:H_sq}}
    H_{01}^{sq}(z, t) = & \Omega e^{-ikz}\sum_{i = 1}^{n} \sin{((2i - 1)(\delta_{sq} + \chi))} \\ 
    & + \Omega e^{ikz}\sum_{j = 1}^{n} \sin{((2j - 1)\delta_{sq})}
\end{split}
\end{equation}

\begin{figure}[b]
\includegraphics[width=0.48\textwidth]{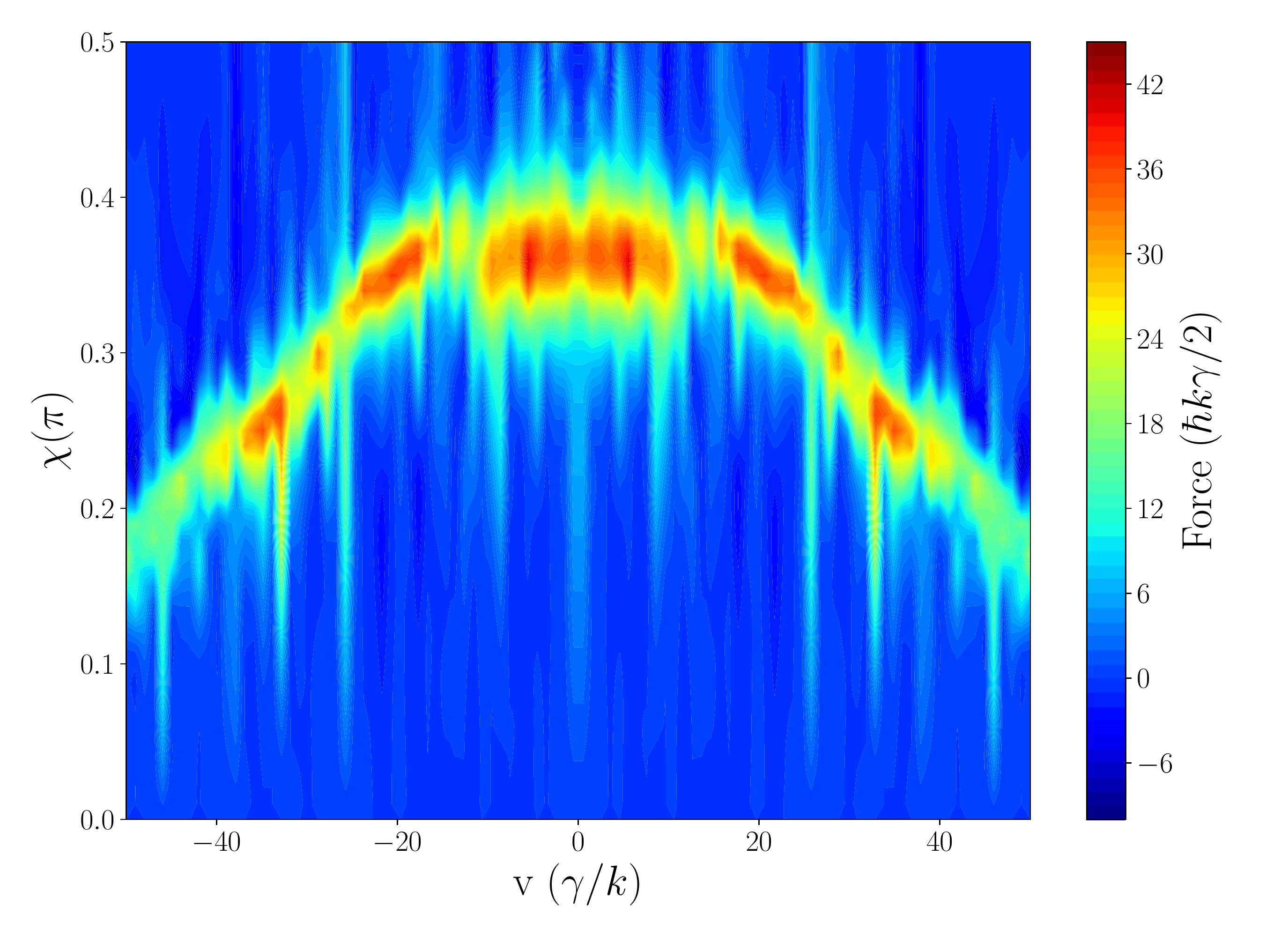}
\caption{\label{fig:FVphase3d} Stimulated force from an approximate square wave modulation with only the first three harmonics ($n = 3$) of $\delta_{sq}$ plotted in the parameter plane of the atom’s velocity $v$ and phase $\chi$. $\Omega = \pi\delta_{sq}/4 = 61 \gamma$, $\chi = 0.36\pi$ produce the strongest stimulated force with a magnitude of over $35F_{rad}$, which acts over a broad velocity range of over $\pm 25 \gamma/k$.}
\end{figure}

Calculating and plotting stimulated forces as a function of velocity over a parameter space spanning a Rabi frequency range $0 \leq \Omega \leq 3\delta_{sq}$ and a phase range $0 \leq \chi \leq \pi$ revealed that $\Omega = \pi \delta_{sq}/4 \approx 0.8\delta_{sq}$ and relative phase $\chi = 0.36\pi \approx \pi/3$ optimized the magnitude and velocity range of the stimulated force. At the optimal Rabi frequency value $\Omega = \pi \delta_{sq}/4$, the force-velocity profile is sensitive to the choice of phase $\chi$. Figure \ref{fig:FVphase3d} shows the contour plot of this square wave force as a function of the atom’s velocity $v$, and phase $\chi$ in the units of $\pi$ between 0 and $\pi/2$ for $\Omega = 61\gamma$. With larger force shown in red, the plot indicates a region between $0.3\pi \leq \chi \leq 0.4\pi$ where a strong stimulated force is uniformly present over a broad velocity range of $\pm 25\gamma/k$. This width is almost twice the width at the peak of the optimal BCF force-velocity profile in Fig.\ref{fig:forcevelocity} for $\delta=50\gamma$ and the same Rabi frequency $\Omega=61\gamma$. For smaller values of $\chi$, the velocity range of the square wave force splits into two narrow domains, and the force vanishes close to $\chi = 0$ and $\pi/2$. We verified from this plot that $\chi = 0.36\pi$ optimized the force magnitude and velocity range. This optimal phase remains the same regardless of the number of harmonics $n$ present in the square wave, so we only display the $n = 3$ case in Fig.\ref{fig:FVphase3d}.

\begin{figure*}[ht]
\includegraphics[width=0.96\textwidth]{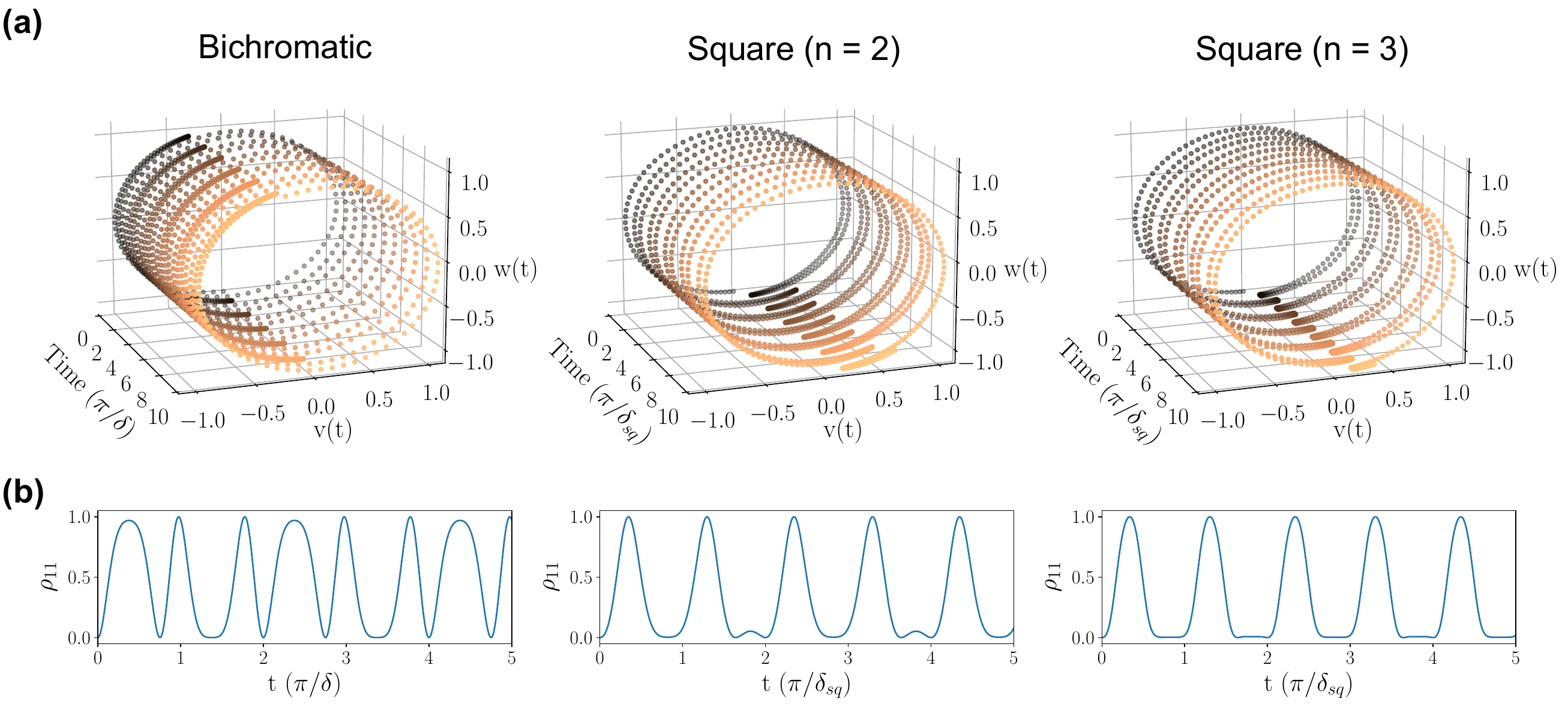}
\caption{\label{fig:Bloch_sq_ns} Time evolution of the Bloch vector (top row) and the excited-state population $\rho_{11}$ (bottom row) for a two-level system in the optimal bichromatic (left) and square wave light fields with the first two harmonics (center), and three harmonics (right).}
\end{figure*}

The optimal $\Omega/\delta_{sq}$ ratio we found matches the $\pi$-pulse condition $\Omega = \pi \delta_{sq}/4$ described in \cite{sodingShortDistanceAtomicBeam1997, williams_bichromatic_2000}.  When this condition is fulfilled, the arrival of each pulse inverts the atomic population between the atom’s ground and excited states. Counter-propagating pulse trains keep the atom cycling between the two states as it absorbs photons from one pulse train and re-emits them into the other. We confirmed this finding at the optimal conditions $\Omega = \pi \delta_{sq}/4, \chi = 0.36\pi$ by plotting the Bloch vector of the two-level system as a function of time. 
The Rabi cylinder plots in Fig.\ref{fig:Bloch_sq_ns}a show the time evolution of two components of the Bloch vector 
\begin{align*}
    v(t) &= i(\rho_{01} - \rho_{10}) \\
    w(t) &= \rho_{11} - \rho_{00}
\end{align*}
without decay for the optimum BCF, and truncated square wave cases with only two and three harmonics $n$ in the square wave Fourier series. The Bloch vector trajectory in the BCF case (left) shows symmetric wrapping around $w(t) = 0$ (note that our result is $\pi/2$ phase-shifted relative to Galica et al.'s result in \cite{galica_four-color_2013}). Starting in the ground state, the Bloch vector rotates clockwise and approaches the excited state, at which it reverses direction, then returns to the ground state before rapidly completing a full counterclockwise rotation. The plot of excited-state population $\rho_{11}$ as a function of time directly below the Bloch cylinder shows that the additional wrapping slows down the cycling between the ground and excited state. These results draw a clear distinction between the bichromatic light fields and alternating $\pi$-pulses, which would produce a simple rotation of the Bloch vector without any wrapping.

\begin{figure}[b]
\includegraphics[width=0.48\textwidth]{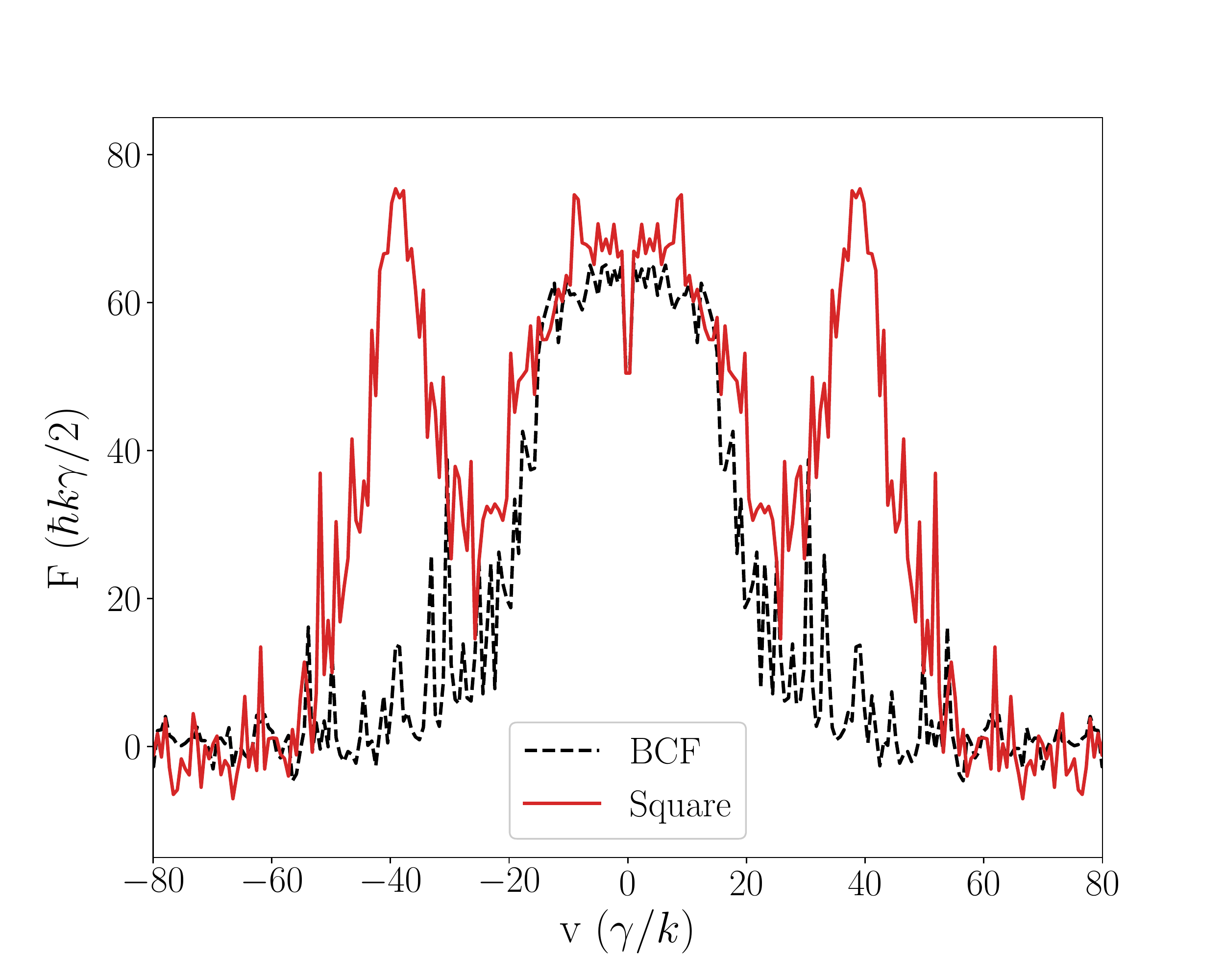}
\caption{\label{fig:FV_BCF_Square} Force-velocity plots showing BCF (black) and the stimulated force from a near-ideal square wave (red). $\Omega = 122\gamma$ for both cases. $\delta = 100\gamma$ in the BCF case to keep $\Omega/\delta = \sqrt{3/2}$ and $\chi=\pi/4$. For the square wave case, $\delta = 155\gamma$ and $\chi = 0.36\pi$ to optimize the force profile.}
\end{figure}

On the other hand, the Bloch vector trajectories under the optimal square wave conditions are similar to the $\pi$-pulse-like periodic cycling between the ground and the excited state with minimal wrapping around the ground state. Figure \ref{fig:Bloch_sq_ns}b shows that on average, an atom spends less time in the excited state when driven by the optimal square wave light fields. This reduces the number of spontaneous decays by 33\% compared to the BCF method, and hence decreases the heating effects from momentum diffusion. The center plot in Fig.\ref{fig:Bloch_sq_ns}a with only two harmonics ($n = 2$)  is very similar to the four-color case presented in \cite{galica_four-color_2013}, which also contains only the first and third harmonics of the bichromatic light fields. As Fig.\ref{fig:Bloch_sq_ns}a shows, adding the 5th harmonic ($n = 3$) suppressed the wrapping around the ground state almost entirely. The excited-state population plot below for $n = 3$ shows a simple oscillation between the ground and excited state, which is more similar to the $\pi$-pulse behavior. When even higher harmonics are present, the Bloch vector simply reverses the direction at the ground state after each complete rotation with virtually no wrapping at the ground state. Although the Bloch vector trajectories slightly differ for different number of harmonics $n$, we found that the force-velocity profile of the square wave force, shown as the solid curve in Fig.\ref{fig:FV_BCF_Square}, remained the same for all $n > 2$ for $\Omega = \pi\delta_{sq}/4$ and $\chi=0.36\pi$. Note that both the strong force and the population inversion disappear if we remove the third and all higher harmonics and revert to a sinusoidal modulation ($n = 1$) under the same conditions.

Figure \ref{fig:FV_BCF_Square} shows that the square wave modulation enhances the velocity range of the stimulated force. The square wave force (red solid curve), generated using the near-ideal square wave, has a comparable maximum magnitude comparable to the BCF of the same optical power (black dashed curve), and consists of three prominent broad peaks instead of one. The two side peaks are symmetrical around zero velocity with about half the width of the central peak, such that the total width is roughly twice that of the BCF profile. As previously stated, the square wave force profile did not change significantly with the number of harmonics $n$ present. The $n = 2$ force profile overlapped almost perfectly with the force profile shown in Fig.\ref{fig:FV_BCF_Square}, which was generated by the sharp-edge square wave function from the \texttt{Scipy} library.

\begin{figure}[ht]
\includegraphics[width=0.48\textwidth]{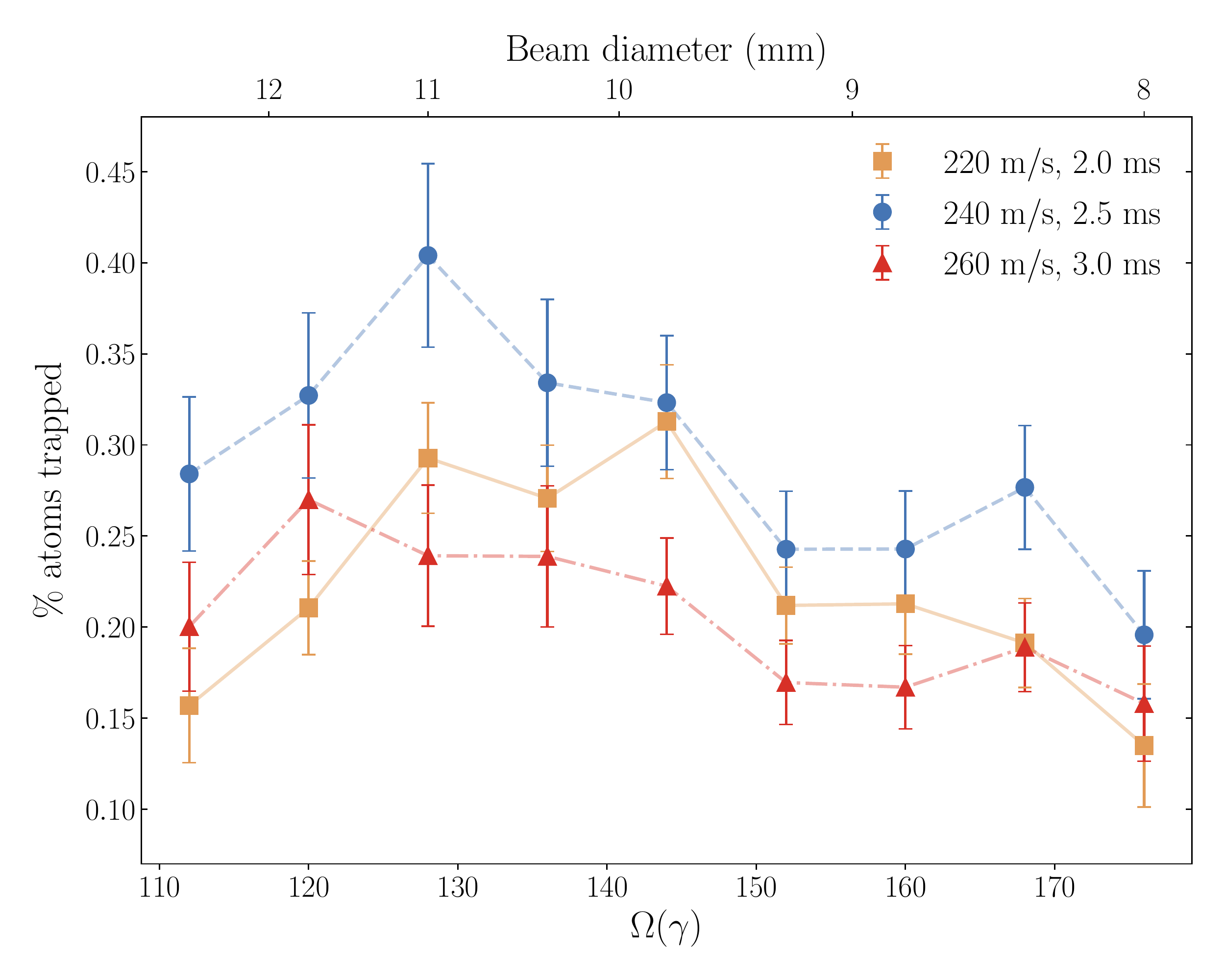}
\caption{\label{fig:contour_square} Percentage of atoms in the atomic beam that can be trapped by the 556 nm MOT after slowing with the square wave stimulated force plotted against Rabi frequency $\Omega$ and laser beam diameter for three different chirp settings: 220 $\rightarrow$ 10 m/s in 2 ms (square, solid line), 240 $\rightarrow$ 10 m/s in 2.5 ms (circle, dashed line), 260 $\rightarrow$ 10 m/s in 3 ms (triangle, dot-dashed line). Up to 0.40\% of atoms can be trapped using $\delta_{sq} = 160\gamma$, $\Omega = 128\gamma$ with a Gaussian beam 1/e diameter of 11 mm, total laser power of 1 W, a slowing distance of 33 cm, and a laser detuning chirp from 240 to 10 m/s in 2.5 ms.}
\end{figure}

After characterizing and optimizing the square wave force profile, we ran Monte Carlo simulations to predict a MOT loading rate under the same experimental conditions we used with the BCF model. After exploring the parameter space of Rabi frequency, laser frequency chirp velocity range, chirp rate, and Gaussian beam size while keeping the total optical power at 1 W, we found that up to 0.40$\pm$0.05\%  of Yb atoms in the atomic beam can be trapped by the 556 nm MOT after being slowed by the square wave stimulated force. Figure \ref{fig:contour_square} shows the predicted loading rate as a function of Rabi frequency $\Omega$ and beam diameter for three different chirp settings: 220 $\rightarrow$ 10 m/s in 2 ms (square, solid line), 240 $\rightarrow$ 10 m/s in 2.5 ms (circle, dashed line), 260 $\rightarrow$ 10 m/s in 3 ms (triangle, dot-dashed line). The optimal conditions favor a larger 1/e Gaussian laser beam diameter of 11 mm compared with the BCF preferred beam diameter of 7.6 mm. Given that our atomic beam has a diameter of only 4 mm, this result highlights the pronounced effects of the Rabi frequency variation in the Gaussian beam on the stimulated force produced by a square wave modulation.

Compared to the chirped BCF method that can produce the MOT loading rate of $2.4 \times 10^8$ atoms/s, our simulation predicts that the square wave method can enhance the loading rate by 70\% ($4.0 \times 10^8$ atoms/s) using the same laser power and a larger Gaussian beam diameter. For the square wave method, we have assumed that electro-optic modulators can produce the square wave amplitude modulation on the two counter-propagating beams at the desired modulation frequency around $\delta_{sq} = 160\gamma \approx 2\pi \times 29$ MHz without losing any optical power, as opposed to splitting the laser beam into multiple laser beams and overlapping the different harmonics. This assumption allowed us to use larger beam sizes in the square wave case, in which 1 W of laser power was split evenly between two beams instead of four beams in the BCF case. We have shown that both the square wave and BCF methods can produce a loading rate on the order of 10\textsuperscript{8} atoms/s using only one 556 nm laser. The MOT loading rate might be improved further if one modifies the laser beam profile from Gaussian to flat-top over the slowing distance between the oven aperture and the MOT. This is difficult in our experiment due to the long slowing distance of 33 cm.

\subsection{\label{subsec:phasemod}Phase modulation}
Both the bichromatic and polychromatic stimulated slowing methods presented so far utilize counter-propagating amplitude-modulated light to induce stimulated emissions in atoms. In this section, we investigate a different approach using phase-modulated light.

\begin{figure}[b]
\includegraphics[width=0.48\textwidth]{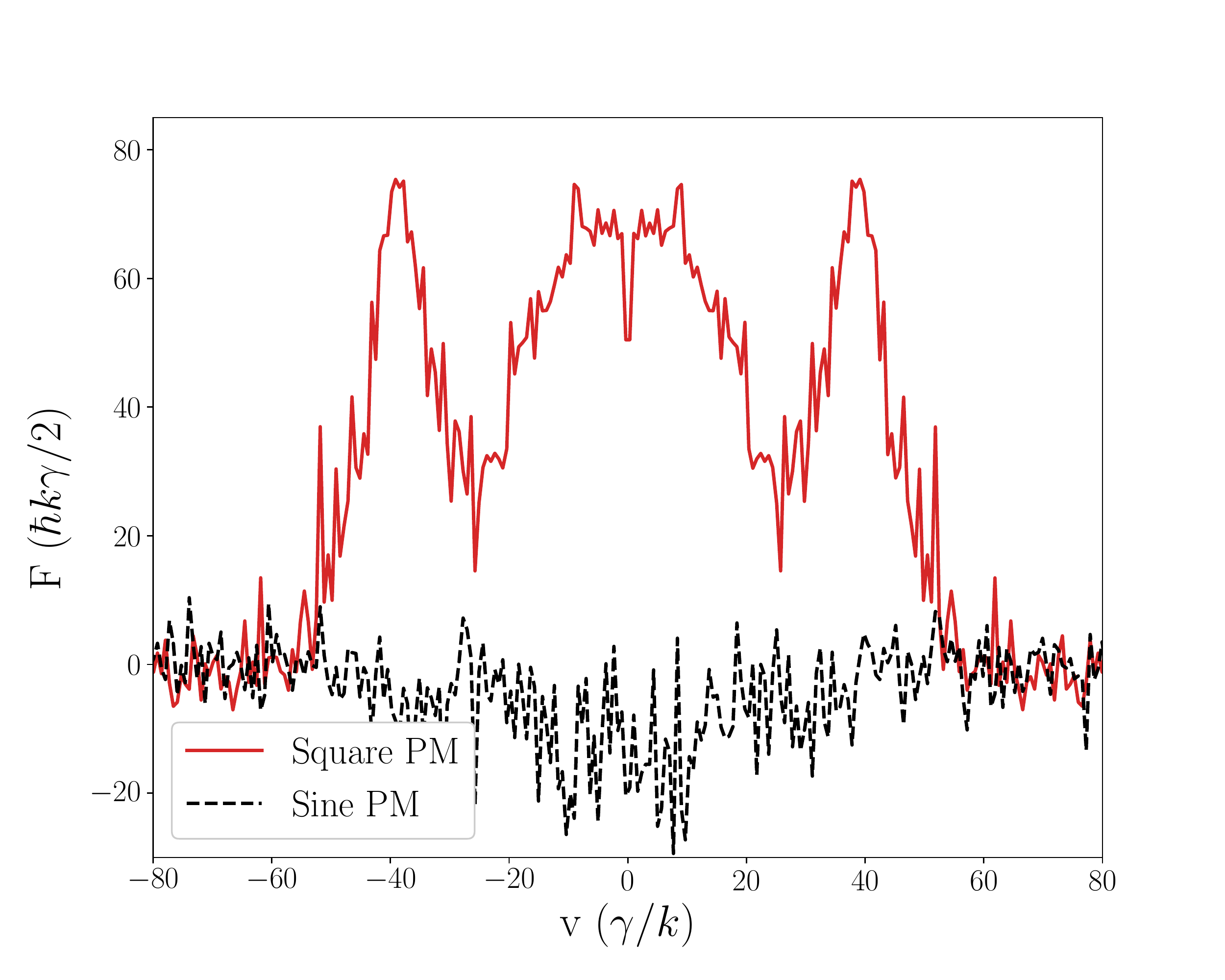}
\caption{\label{fig:phasemod_test} Force-velocity plots for square wave phase modulation function (solid) and sinusoidal phase modulation function (dashed). $\delta_{\phi} = 153\gamma,\Omega=122\gamma$, and $\chi=0.36\pi$ in both cases.}
\end{figure}

Motivated by the promising results from Sec.\ref{subsec:square}, we first looked at a square wave phase modulation (square wave PM). In the time domain, a square wave amplitude modulation (square wave AM) that alternates the sign of an envelope between 1 and -1 produces an equivalent output as a square wave PM that alternates the optical phase between 0 and $\pi$. We verified with our model that a square wave PM can produce an identical stimulated force-velocity profile (solid curve in Fig.\ref{fig:phasemod_test}) to the square wave AM force profile shown in Fig.\ref{fig:FV_BCF_Square}, given that the phase modulation frequency $\delta_{\phi}$ and Rabi frequency $\Omega$ satisfy $\Omega = \pi\delta_{\phi}/4$ and the relative phase between the two counter-propagating beams is fixed at $\chi = 0.36\pi$. This result suggests that it is possible to replace amplitude modulators by phase modulators in a stimulated slowing experiment to produce the same force by a square wave modulation. As with the square wave AM, replacing the near-ideal square wave function with a truncated Fourier series did not alter the force-velocity profile shown in Fig.\ref{fig:phasemod_test}. 

We then tried replacing the square wave modulation with a sinusoidal phase modulation (sine PM) while keeping $\pi\delta_{\phi}/4$ and $\chi = 0.36\pi$. As shown in Fig.\ref{fig:phasemod_test}, the force (dashed curve) almost vanishes and no longer looks symmetrical about zero velocity. In the time domain, beside the larger wrapping of the Bloch vector near the ground state, we did not identify any striking features that could explain this asymmetry in the force-velocity profile. We have also tried searching the parameter space of $\Omega/\delta_{\phi}$ and $\chi$ to identify new optimizing conditions for the sinusoidal PM setting, but we have not found any conditions that the sinusoidal phase modulation can improve the stimulated force beyond the bichromatic force, or the square wave force.

\subsection{\label{subsec:broadband}Broadband cooling}
To broaden the velocity range of the BCF and enhance the MOT loading rate, we also investigated a broadband cooling approach by adding bichromatic light fields at different velocity detunings.  This is similar to ideas of broadband cooling that have proven useful in some cases of Doppler cooling \cite{hoffnagle1988proposal, wallis1989broadband, zhu1991continuous, watanabe1996velocity, loftus2004narrow}. However, for our configuration of stimulated slowing on a two-level atom, interference can make the force vanish if the two velocity detunings are less than $\sim2\delta/k$ apart.

\begin{figure}[b]
\includegraphics[width=0.48\textwidth]{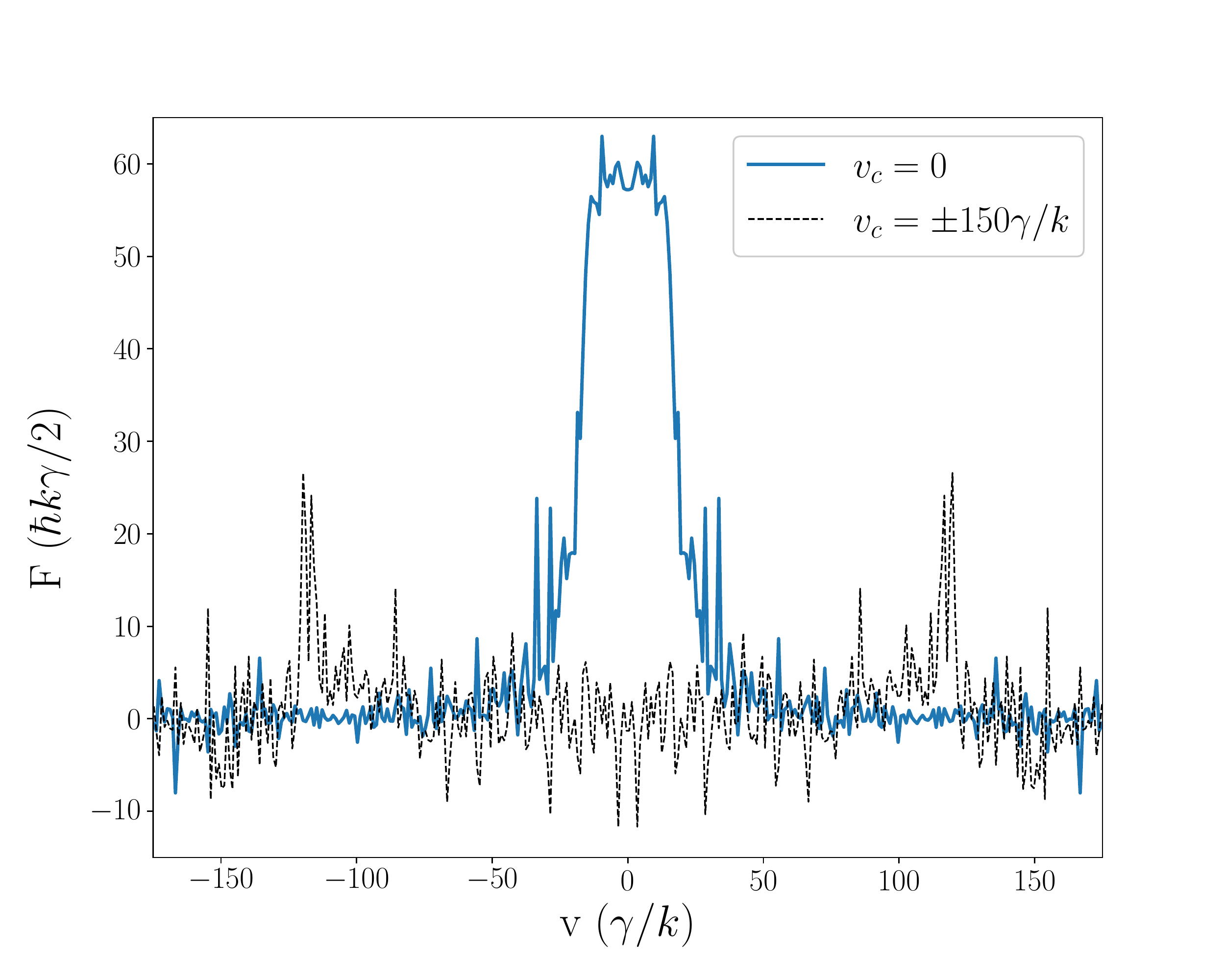}
\caption{\label{fig:broadband_test} Bichromatic force-velocity plots centered at zero detuning $v_c = 0$ (solid) and two velocity detunings at $v_c =\pm 1.5\delta/k = \pm150\gamma/k$ (dashed).}
\end{figure}

Some stimulated forces can be retrieved when the velocity detunings are further apart. Figure \ref{fig:broadband_test} shows the bichromatic force-velocity plots at zero velocity detuning $v_c = 0$ (solid) and at two velocity detunings $v_c =\pm 1.5\delta/k = \pm150\gamma/k$ (dashed). Two bichromatic light fields of the same $\delta = 100\gamma$ and $\Omega = \sqrt{3/2}\delta$ at two detunings separated by $3\delta/k$ produce two small peaks that are much smaller and narrower than the original force profile. The minimum spacing required $\sim 2\delta/k$ to retrieve the forces is greater than the width of an individual BCF profile $\Delta v \sim \delta/2k$. Therefore, we cannot produce a broadened stimulated force on a single two-level system by simply adding more detunings at the same phase $\chi=\pi/4$.

We also verified that it was not possible to produce optical molasses on a single two-level system by combining two shifted bichromatic force profiles at the opposite phases $\chi=\pm\pi/4$. This is consistent with Partlow et al.’s findings \cite{partlow2004bichromatic} from their attempt to create the optical molasses force on He* atoms. To avoid interference, they had to apply the two force profiles on the atoms at two separate locations. On the other hand, as demonstrated by Wenz et al., in more complicated systems such as diatomic and polyatomic radicals that can be represented by two coupled two-level systems, large molasses-like cooling forces can be realized using two separated BCF profiles \cite{wenz_large_2020}. This suggests that broadband cooling, which combines multiple BCF profiles of the same phase at different detunings, may be feasible in more complicated systems than a single two-level atom. Overall, the most effective method we have found to extend the velocity range of the stimulated forces on two-level atoms is the laser frequency chirp method \cite{chiedaBichromaticSlowingMetastable2012}, which can be applied to either bichromatic and polychromatic forces.

\section{\label{sec:conclusion}Conclusion}
We developed a numerical model and Monte Carlo simulations to analyze bichromatic and polychromatic stimulated forces for slowing of Yb atoms under realistic experimental conditions. We have shown theoretically that it is possible to cool and trap Yb atoms on the narrow \textsuperscript{3}P\textsubscript{1} transition and achieve a MOT loading rate in the order of 10\textsuperscript{8} atoms/s using the total laser power of 1 W (Rabi frequency $\Omega$ near 25 MHz).  

The square wave modulation results show great promise. At optimal conditions, this polychromatic force has about the same magnitude but almost twice the velocity range of the bichromatic force. Adding higher harmonics to the square wave modulation makes the pulses more similar to $\pi$-pulses as shown in the Bloch vector trajectories. This square-wave stimulated force may be realized by modulating either the amplitude or phase of two counter-propagating beams, instead of overlapping four bichromatic CW beams. Splitting total laser power evenly between two beams instead of four partially accounts for the 70\% enhancement in the predicted MOT loading rate compared to the chirped BCF method. Our simulation also shows that the square wave modulation reduces spontaneous emissions by up to 33\% compared to the BCF method. This can help prevent heating, especially in atoms with shorter lifetimes, and population loss in multilevel systems and molecules. 

Our model has informed optimal experimental design for stimulated slowing and trapping of Yb atoms. Experiments are underway to test and verify these findings. The simulations we developed have flexibility and can be modified to investigate other two-level systems, such as other alkaline-earth-line atoms (e.g. Ca, Sr).

\begin{acknowledgments}
This research was funded by Office of Naval Research grant N000141712255. Some of the computing was performed on the Sherlock cluster at Stanford University. We would like to thank the Stanford Research Computing Center for providing computational resources and support.
\end{acknowledgments}


\bibliography{BBT_refs.bib}

\end{document}